%% file: ClanIII.tex
\documentstyle{ioplppt}
\begin{document}
\eqnobysec
\input makra1
\jl{6}
\title{Curvature invariants in type-{\it{III}} spacetimes}
\author{V. Pravda{\ftnote{4}{E-mail: pravda@math.cas.cz}}}
\address{Mathematical Institute, 
Academy of Sciences, \protect\\
\v Zitn\' a 25, 115 67 Prague 1, Czech Republic}

\begin{abstract}
The results of paper \cite{BicVoj} are generalized for  vacuum type-{\it{III}} solutions
with, in general, a non-vanishing cosmological constant $\Lambda$.
It is shown that all curvature invariants containing derivatives of the Weyl tensor vanish if a 
type-{\it{III}} spacetime admits a non-expanding and non-twisting null geodesic congruence.

A non-vanishing curvature invariant containing first derivatives of the Weyl tensor 
is found in the case of type-{\it{III}} spacetime with expansion or twist.
\end{abstract}
\pacs{0420, 0430}

\section{Introduction}

In \cite{BicVoj} we proved  that in Petrov type-{\it N}  vacuum
spacetimes which admit  a non-expanding  and non-twisting
null geodesic congruence all curvature invariants constructed
from the Weyl tensor and its  derivatives of arbitrary order vanish. 
We generalize this paper and  obtain the same result for
non-expanding  and non-twisting
Petrov type-{\it III}  vacuum spacetimes. 
Thus it is useful to study these 
spacetimes  in quantum gravity, as  all their
quantum corrections vanish (see Gibbons \cite{Gibbons}).
The proof for
 type-{\it III}  vacuum spacetimes is based on the same ideas as
that  for type-{\it N}  vacuum spacetimes given in \cite{BicVoj}. 
Here we just outline the basic ideas of the proof 
(see Section 3). For understanding and rigorous reconstruction of the proof, paper \cite{BicVoj} is indispensable.

In the case of type-{\it III} vacuum spacetime with expansion or twist
we find a nonzero curvature invariant of the first order (containing 
the first derivatives of the Weyl tensor). 

First let us recall some basic relations from spinor calculus 
and Newman-Penrose formalism.
We can use basis
 {$\sod{A} $, $\sid{A} $}, which satisfies
\BE
\sod{A} \siu{A} =1,\quad \sod{A} \sou{A} =0, \quad \sid{A} \siu{A} =0, \label{basis}
\EE
 to decompose the Weyl spinor (see \cite{PR})
\BEA
\ospd{\Psi}{ABCD} =&& \Psi_0 \sid{A} \sid{B} \sid{C} \sid{D} 
                                     - 4 \Psi_1 \sod{(A} \sid{B} \sid{C} \sid{D)}
                                     + 6 \Psi_2 \sod{(A} \sod{B} \sid{C} \sid{D)} \nonumber \\
                                    &&-  4 \Psi_3 \sod{(A} \sod{B} \sod{C} \sid{D)}
                                     + \Psi_4  \sod{A} \sod{B} \sod{C} \sod{D}  \label{gendec}   \ ,
\EEA
where
\BEA
\Psi_0 &=&  \ospd{\Psi}{ABCD} \sou{A} \sou{B} \sou{C} \sou{D} \nonumber \ ,  \\
\Psi_1 &=&  \ospd{\Psi}{ABCD} \sou{A} \sou{B} \sou{C} \siu{D}  \nonumber \ ,   \\
\Psi_2 &=&  \ospd{\Psi}{ABCD} \sou{A} \sou{B} \siu{C} \siu{D} \ ,  \\
\Psi_3 &=&  \ospd{\Psi}{ABCD} \sou{A} \siu{B} \siu{C} \siu{D}  \nonumber  \ ,  \\
\Psi_4 &=&  \ospd{\Psi}{ABCD} \siu{A} \siu{B} \siu{C} \siu{D} \ .     \nonumber 
\EEA
There exist four principal spinors $\ospd{\alpha}{A} $, $\ospd{\beta}{A} $,  $\ospd{\gamma}{A} $, $\ospd{\delta}{A} $ such that 
\BE
\ospd{\Psi}{ABCD} = \ospd{\alpha}{(A} \ospd{\beta}{B} \ospd{\gamma}{C}  
 \ospd{\delta}{D)}  \ .
\EE
Since  three principal spinors of $\ospd{\Psi}{ABCD} $  coincide  in type-{\it{III}} spacetimes,  it is convenient to choose this repeated principal spinor as a basis spinor $\sod{A} $. Then
\BE
\ospd{\Psi}{ABCD} = \sod{(A} \sod{B} \sod{C}  \ospd{\delta}{D)} 
\EE 
and
\BE
\Psi_0 = \Psi_1 =\Psi_2 = 0 \ . \label{NPV0}
\EE
For the Weyl spinor we thus get
\BE
\ospd{\Psi}{ABCD} =
                                    -  4 \Psi_3 \sod{(A} \sod{B} \sod{C} \sid{D)}
                                     + \Psi_4  \sod{A} \sod{B} \sod{C} \sod{D}  \ .   \label{rozkweyl3}
\EE
We choose the second basis spinor $\sid{A} $ to satisfy
\BE
D \sid{A} = 0\ ,
\EE
which implies  that 
a complex null tetrad induced by $\sod{A} $ and $\sid{A} $
is parallelly propagated along the geodetic null congruence
and several Newman-Penrose coefficients vanish:
\BE
\sigma = \kappa = \ne = \pi = 0.       \label{NPC0}
\EE

To  end  this section let us write down the relations
\BE
\dsu{A}{X} = \siu{A} \csiu{X} D + \sou{A} \csou{X} \T - \siu{A}
\csou{X} \d - \sou{A} \csiu{X} \cd   \ ,   \label{rozds}
\EE
\BEA
 \dsu{A}{X} \sou{B} &=&\ng \sou{A} \sou{B} \csou{X}
                     -\na   \sou{A} \sou{B} \csiu{X}
                     -\nt   \sou{A} \siu{B} \csou{X}
                     -\nb   \siu{A} \sou{B} \csou{X} \nonumber \\
                     &+& \nr   \sou{A} \siu{B} \csiu{X}
                     +\ne   \siu{A} \sou{B} \csiu{X}
                     +\ns   \siu{A} \siu{B} \csou{X}
                     -\nk   \siu{A} \siu{B} \csiu{X}   \ ,  \label{rozdso}\\
 \dsu{A}{X} \siu{B} &=& \nn  \sou{A} \sou{B} \csou{X}
                     -\nl   \sou{A} \sou{B} \csiu{X}
                     -\ng   \sou{A} \siu{B} \csou{X}
                     -\nm   \siu{A} \sou{B} \csou{X} \nonumber \\
                     &+&\na   \sou{A} \siu{B} \csiu{X}
                     +\np   \siu{A} \sou{B} \csiu{X}
                     +\nb   \siu{A} \siu{B} \csou{X}
                     -\ne   \siu{A} \siu{B} \csiu{X} \ .   \label{rozdsi}
\EEA

Equations  (\ref{basis}) and (\ref{rozkweyl3}) imply that all invariant quantities
constructed from $\ospd{\Psi}{ABCD} $ without derivatives vanish and thus
all curvature invariants of the zeroth order vanish too. In next sections we
study curvature invariants of higher orders.

\section{Expanding or twisting solutions}

Regarding (\ref{NPV0}) and (\ref{NPC0}), the Bianchi identity (see Eq. (7.67) in \cite{Kramer})
gives 
\BE
D \Psi_3 = 2  \rho \Psi_3\ . \label{Bid1}
\EE
Using (\ref{basis}), (\ref{rozds})  -  (\ref{rozdsi}), one can easily show that all first order invariants
of the   Weyl tensor vanish if they contain  only squares or cubes in $C_{\na \nb \ng \d ; \ne }$.
However, there is a non-vanishing curvature invariant
\BE
I \ = \ C^{\na \nb \ng \d ; \ne } C_{\na \nm \ng \nn ; \ne } C^{\nl \nm \nr \nn ; \ns }
C_{\nl\nb\nr\d;\ns }\ ,     \label{tvarinvC}
\EE
which, in terms of Newman-Penrose quantities, reads
\BE
I = (48 \rho \bar \rho \Psi_3 \bar  \Psi_3)^2. \label{INP}
\EE
The Robinson-Trautman metric of type-{\it{III}}, that is the general vacuum type-{\it{III}} solution 
admitting a geodesic, shearfree, twistfree and diverging null congruence,
has the form
\BE
{\rm d}s^2 = \frac{2r^2}{P^2}{\rm d} \zeta {\rm d} \bar \zeta -2 {\rm d} u {\rm d} r - ( \T \ln P - 2r {(\ln P),}_u )  
{\rm d} u^2 \ ,
\EE
where $P(u,\zeta, \bar \zeta)$ satisfies 
\BE
\T \T P = 0\ , \quad  {(\T \ln P),}_{\zeta} \not= 0 \ , \quad  \T \equiv 2 P^2 \partial_{\zeta} \partial_{\bar \zeta} 
\EE
and $\partial/ \partial r$ is the repeated null eigenvector.
In an appropriately chosen complex null tetrad (given for example in Chapter 23 in \cite{Kramer} )
we obtain 
\BEA
\ns = \nk = \ne = \np = \Psi_0 = \Psi_1 = \Psi_2 = 0 \ , \quad   \nr = -\frac{1}{r} \nonumber \ , \\
\Psi_3 =  -\frac{P}{r^2}{(\T \ln P),}_{\bar \zeta}  \label{RTN}  \ , \\ 
\Psi_4 = \frac{1}{r^2} {\left( P^2 {\left( \frac{1}{2}  \T \ln P - r {(\ln P),}_u \right)}_{, \bar \zeta} \right)}_{, \bar \zeta} \nonumber \ .
\EEA
Substituting (\ref{RTN}) into the invariant (\ref{INP}), we get
\BE
I =  \left(\frac{48}{r^6} P {\bar P} {(\T \ln P),}_{\bar \zeta}  {(\T \ln \bar P),}_{ \zeta} \right)^2 \ .
\EE
This  invariant, which is non-zero in general, can be used  for analyzing singularities
in Robinson-Trautman solutions.

\section{Non-expanding and non-twisting solutions}

Non-expanding and non-twisting solutions satisfying (\ref{NPC0})
and $\nr=0$ belong to Kundt's class and they are completely known 
(see Chapter 27. 5. 1. in \cite{Kramer}).

In \cite{BicVoj} we have proved that for type-{\it N} vacuum spacetimes, without expansion and without twist,  all curvature invariants of all orders vanish.
This proof, with  slight modifications, is also valid  for type-{\it{III}} vacuum spacetimes
without expansion and without twist. Thus  we  give here
only  the basic ideas of the proof.

In the following we need
NP equations containing operator $D$: 
\BEA
D \tau = 0 \nonumber \ ,  \\
D \alpha = 0  \nonumber \ , \\
D \beta = 0   \label{NPeq} \ , \\
D \gamma = \nt \na + \nct \nb - R/24 \ , \nonumber \\
D \nl = 0  \nonumber \ , \\
D \nm = R/12  \nonumber \ , \\
D \nn = \nct \nm + \nt \nl + \Psi_3 \ ,  \nonumber
\EEA 
and the commutators
\BEA
(\T D - D\T) &=&  (\ng+\ncg)D-\nt\cd - \nct\d\
 \nonumber ,\\
(\d D - D\d) &=&  (\nca+\nb)D\ . \label{com}
\EEA
The Bianchi identity (\ref{Bid1}) has the form
\BE
D \Psi_3 = 0 \ . \label{DP3}
\EE

Let us now turn attention to the behaviour of the NP quantities  under the constant boost
transformation
\BE
\ospu{o'}{A} = a \sou{A} \ , \quad
 \ospu{\iota'}{A} = {a}^{-1} \siu{A} \ .
 \label{boost}
\EE
A quantity $\Omega$,  which transforms under this boost as
\BE
\Omega'=a^q \Omega\  , \label{trOmg}
\EE
has the boost-weight $b(\Omega)=q$.
Summary of the boost-weights for NP coefficients (NP) and operators (OP) is given
in Table 1 in \cite{BicVoj}.
For  $\Psi_3 $ we have
\BE
{\Psi'}_3=a^{-2} \Psi_3  \Longrightarrow
  b({\Psi_3})= -2\ .    \label{bPsi3}
\EE

Now we analyze invariants of $\nderPt$.
The quantity  $\Psi_3 \sou{(A} \sou{B} \sou{C} \siu{D)} $ is invariant under the boost transformation
(\ref{boost})
\BE
{\Psi'}_3 \ospu{o'}{(A} \ospu{o'}{B}  \ospu{o'}{C} \ospu{\iota'}{D)} = 
{\Psi}_3 \ospu{o}{(A} \ospu{o}{B}  \ospu{o}{C} \ospu{\iota}{D)} 
\EE
and thus also $\nderPt$ is invariant under (\ref{boost}) and $b(\nderPt)=0$. 
Using Leibniz's formula and relations (\ref{rozds})-(\ref{rozdsi}),  we decompose
the spinor derivative $\nderPt$ into the spinor basis  of the appropriate
spinor space. Each term in such a sum has the form
\BE
K  \underbrace{\sou{A_1} \dots  \sou{A_{m_1} } }_{m_1}
         \underbrace{\csou{X_1} \dots  \csou{X_{m_2} } }_{m_2}
         \underbrace{\siu{B_1} \dots  \siu{B_{n_1} } }_{n_1}
         \underbrace{\csiu{Y_1} \dots  \csiu{Y_{n_2} } }_{n_2} \ ,
\EE
where $K$ is a product of NP quantities. 
This term is also invariant under  the boost (\ref{boost})
and thus
\BE
b(K) = n_1 + n_2 - m_1 - m_2 \ .
\EE
In the following
we show that NP equations imply  $K=0$ if $b(K) \geq 0$ and thus
the decomposition of $\nderPt$ consists 
only of terms containing more $o$'s  then $\iota$'s and,  as a consequence of
Eq. (\ref{basis}), all invariants of  $\nderPt$ vanish. \\
{\bf Lemma 1:} \\
Let an invariant constructed from the products of the  spinors \\
$\nderPt$, for fixed $n$,  be non-vanishing. Then there exists a
non-vanishing quantity $K=X_1 X_2 \dots X_n \Psi_3 $,
$X_i \in NP \cup OP $   such that
\BDM b(X_1 X_2 \dots X_n \Psi_3 ) =
\sum_{i=1}^{n}  b(X_i)  + b(\Psi_3) \geq 0 , {\mbox { i.e. }}, \ \ 
\sum_{i=1}^{n}  b(X_i)  \geq  2\ .
\EDM
See the proof in \cite{BicVoj}. 

We introduce a number $p$ for each
NP-coefficient (or its derivatives) that describes its behaviour
under the action of the operator $D$ (see \cite{BicVoj} for the
exact definition and Table 2).

Comparing (\ref{NPeq}) and (\ref{com}) with Eqs. (3.2)-(3.4) in \cite{BicVoj}
we see that Table 2 and Lemma 2 in \cite{BicVoj} remain unchanged.
This  enables us to reformulate Lemma 3 (as a consequence
of  (\ref{NPeq})-(\ref{DP3})) and Proposition 1: \\
{\bf Lemma 3:} \\
Consider a quantity
$ X_1 X_2 \dots X_n $ where $X_i \in NP \cup OP $.
\BDM
\mbox{If } \sum_{i=1}^{n} p (X_i) < 0  \mbox{ then }
  X_1 X_2 \dots X_n \Psi_3 =0\ .
\EDM
The proof of  Lemma 3 remains unchanged. \\
{\bf Proposition 1:} \\
In type-{\it III} vacuum spacetimes with $\Lambda$ admitting a non-expanding
and non-twisting null geodesic congruence all $n$-th order invariants
formed from the products of spinors $\nderPt$, with $n$ arbitrary
but fixed, vanish.\\[1mm]
Proof:
We follow a similar procedure as in  \cite{BicVoj}, replacing 
 Eq. (3.8) in \cite{BicVoj}
by  
\BE \sum_{i=1}^{n} b (X_i) \geq 2  \ \ \mbox{and} \ \
\sum_{i=1}^{n} p (X_i) \geq 0  \ ,  \label{eqProof}
\EE
which  leads  to the same conclusion.

Invariants constructed from the second term in (\ref{rozkweyl3})  and its derivatives,
 $\nderPsi $, also vanish. The proof is similar to that in  \cite{BicVoj},
with the only difference 
\BEA
D \Psi_4 &=& 0 \quad \mbox{for type-{\it N}} \ , \\
D \Psi_4 &=& (\cd-2 \na) \Psi_3 \ ,  \quad D^2 \Psi_4=0 \quad \mbox{for type-{\it{III}}} \ ,
\EEA
and thus we reformulate Lemma 3 of \cite{BicVoj}: \\
{\bf Lemma 3':} \\
Consider  a quantity
$ X_1 X_2 \dots X_n $ where $X_i \in NP \cup OP $.
\BDM
\mbox{If } \sum_{i=1}^{n} p (X_i) < -1  \mbox{ then }
  X_1 X_2 \dots X_n \Psi_4 =0\ .
\EDM 
Proposition 1 remains unchanged: \\
{\bf Proposition 1':} \\
In type-{\it III} vacuum spacetimes with $\Lambda$ admitting a non-expanding
and non-twisting null geodesic congruence all $n$-th order invariants
formed from the products of spinors $\nderPsi$, with $n$ arbitrary
but fixed, vanish.\\[1mm]

Since each term in decompositions of the spinors $\nderPsi$ and $\nderPt$ into the spinor basis  contains more $o$'s then $\iota$'s, we conclude our analysis with
the following propositions: \\
{\bf Proposition 3:} \\
In type-{\it III} vacuum spacetimes with $\Lambda$, admitting a non-expanding
and non-twisting null geodesic congruence,
all invariants constructed from $\ospd{\Psi}{ABCD} $, 
$\ospd{{\bar{\Psi}}}{\dot A\dot B\dot C\dot D} $
and   their  derivatives of arbitrary  order vanish. \\[1mm]
The same proposition formulated  in the tensor formalism reads:\\
{\bf Proposition 4:} \\
In type-{\it III} vacuum spacetimes with $\Lambda$, admitting a
non-expanding and non-twisting null geodesic congruence,
all invariants constructed  from the Weyl tensor and its
covariant derivatives of arbitrary order vanish. 

\section{Conclusion}

A general Petrov type spacetime has non-zero curvature invariants
of the zeroth order. For type-{\it N} spacetimes we have shown in \cite{BicVoj}
that  all Weyl's invariants of all orders for non-twisting and non-expanding solutions vanish. For twisting or expanding solutions  of type-{\it N}
we have proven that Weyl's invariants of the zeroth and  first orders
vanish but we have found a non-zero invariant of the second order
\BE
 C^{\na \nb \ng \d ; \ne \phi} C_{\na \nm \ng \nn ; \ne \phi} C^{\nl \nm \nr \nn ; \ns \nt}
C_{\nl\nb\nr\d;\ns\nt} =( 48 \nr^2 \ncr^2 {\Psi_4} {\bar \Psi_4} )^2 .
\EE.

In this paper we show that for type-{\it{III}} vacuum spacetimes without twist
and expansion all Weyl's invariants of all orders vanish. This fact can be used in quantum gravity
(see \cite{Gibbons}).  In the case
with expansion or twist only invariants of the zeroth order vanish and there
exists  a non-vanishing invariant of the first order
\BE
\ C^{\na \nb \ng \d ; \ne } C_{\na \nm \ng \nn ; \ne } C^{\nl \nm \nr \nn ; \ns }
C_{\nl\nb\nr\d;\ns } = (48 \rho \bar \rho \Psi_3 \bar  \Psi_3)^2 \ . \label{conclinv}
\EE
This  invariant can be used  for analyzing singularities
in   type-{\it{III}} vacuum spacetimes with expansion or twist. 
The form of the invariant in terms of NP quantities (\ref{conclinv})
can be also helpful for constructing approximate solutions of Einstein's vacuum field equations in the  type-{\it{III}} with twist (see \cite{Alevey} for  type-{\it{N}} ).

Let us summarize our  results in  Table 1:

\begin{table}{{\bf Table 1}: Curvature invariants in  vacuum spacetimes (0 - vanish, 1 - do not vanish)  }\\[1ex]
  \begin{tabular}{|c|c|c|c|c|c|}\hline
     Petrov type                                        &  I, II, D    & III     & III     &  N    & N     \\ \hline
     expansion and twist                          &     & $\nr \not= 0$     & $\nr =0$     & $\nr \not =0$     & $\nr =0 $     \\ \hline
     curvature invariants of order 0       &    1  & 0     & 0     &  0    & 0     \\ \hline
     curvature invariants of order 1       &     1  &  1   & 0     &  0    & 0     \\ \hline
     curvature invariants of order 2       &     1  &  1    & 0     &  1    & 0     \\ \hline
     curvature invariants of order $>2$ &  1    &   1    &  0    &   1   &  0    \\ \hline
  \end{tabular} \\[0.5ex]
{}
\end{table}

\ack
I thank   Alena Pravdov\' a  for useful discussions.
I also acknowledge the supports  from the grant GACR-201/97/0217.

\end{document}

%% file: makra1.tex
%
\def \dcxi {d \bar \xi}
\def \cxi {\bar \xi}
\def \cA {\bar A}
\def \ffi {\varphi}
\def \vs {\longleftrightarrow}
\def \Bar#1 { \overline{#1} }
\def \der {\partial}
\def \cder {\bar \partial}
\def \zt {\zeta}
\def \czt {\bar \zeta}
\def \cL {\bar L}
\def \mbd {{\mbox{d}}}
\def \eps {\varepsilon}
\def \BE {\begin{equation}}
\def \EE {\end{equation}}
\def \BEAH {\begin{eqnarray*}}
\def \EEAH {\end{eqnarray*}}
\def \BEA {\begin{eqnarray}}
\def \EEA {\end{eqnarray}}
\def \BDM {\begin{displaymath}}
\def \EDM {\end{displaymath}}
\def \na {\alpha}
\def \nca {\bar \alpha}
\def \nb {\beta}
\def \ncb {\bar \beta}
\def \ng {\gamma}
\def \ncg {\bar \gamma}
\def \ne {\varepsilon}
\def \nce {\bar \varepsilon}
\def \nk {\kappa}
\def \nck {\bar \kappa}
\def \nl {\lambda}
\def \ncl {\bar \lambda}
\def \nm {\mu}
\def \ncm {\bar \mu}
\def \nn {\nu}
\def \ncn {\bar \nu}
\def \np {\pi}
\def \ncp {\bar \pi}
\def \nr {\rho}
\def \ncr {\bar \rho}
\def \ns {\sigma}
\def \ncs {\bar \sigma}
\def \nt {\tau}
\def \nct  {\bar \tau}
\def \cd {\bar \delta}
\def \d {\delta}
\def \T {\Delta}
\def \cm {\bar m}
\def \dsu#1#2 {\nabla^{ {\mbox{ {\tiny $\!\!\!\! #1 \!\dot #2 $}\rm}}} }
\def \dsd#1#2 {\nabla_{\mbox{{\tiny $\!\!\! #1 \!\dot#2 $}\rm}}}
\def \sou#1 {o^{{\mbox{{\tiny $ #1 $}\rm}}}}
\def \sod#1 {o_{{\mbox{{\tiny $ #1 $}\rm}}}}
\def \csou#1 {{\bar o}^{{\mbox{{\tiny $ \dot #1 $}\rm}}}}
\def \csod#1 {{\bar o}_{{\mbox{{\tiny $ \dot #1 $}\rm}}}}
\def \siu#1 {\iota^{{\mbox{{\tiny $ #1 $}\rm}}}}
\def \sid#1 {\iota_{{\mbox{{\tiny $ #1 $}\rm}}}}
\def \csiu#1 {{\bar \iota}^{{\mbox{{\tiny $ \dot #1 $}\rm}}}}
\def \csid#1 {{\bar \iota}_{{\mbox{{\tiny $ \dot #1 $}\rm}}}}
\def \om#1  { {\bf \omega^{ \!\!\! {\hat {\mbox{ {\tiny #1 } } } }}} \rm}
\def \com#1  { {\bf {\bar \omega}^{ \!\!\! {\hat {\mbox{ {\tiny #1 } } } }}} \rm}

\def \ospu#1#2 {#1^{ {\mbox{ {\tiny $\!\!\! #2 $} \rm}}} }
\def \ospd#1#2 {#1_{ {\mbox{ {\tiny $\!\!\! #2 $} \rm}}} }
\def \ospud#1#2#3 {#1^{ {\mbox{ {\tiny $\!\!\! #2 $} \rm}}}
                     _{ {\mbox{ {\tiny $\!\!\! \ \ #3 $} \rm}}}}
\def \ospdu#1#2#3 {#1_{ {\mbox{ {\tiny $\!\!\!  #2 $} \rm}}}
                     ^{ {\mbox{ {\tiny $\!\!\! \ \ #3 $} \rm}}}}
\def \SSu#1#2  {\mathop{S^{\scriptscriptstyle #2}}
\limits_{\scriptscriptstyle [#1] \hfill}}
\def \SSd#1#2  {\mathop{S_{\scriptscriptstyle #2}}
\limits_{\scriptscriptstyle [#1] \hfill}}
\def \cSSu#1#2  {\mathop{{\bar S}^{\scriptscriptstyle #2}}
\limits_{\scriptscriptstyle [#1] \hfill}}
\def \cSSd#1#2  {\mathop{{\bar S}_{\scriptscriptstyle #2}}
\limits_{\scriptscriptstyle [#1] \hfill}}
\def \Lab {L^{ {\mbox{ {\tiny $\!\!\! A $} \rm}}}
            _{ {\mbox{ {\tiny $     \ B $} \rm}}} }
\def \Su#1#2 {S^{ {\mbox{ {\tiny $\!\!\! #1 $} \rm}}}_{#2}}
\def \Sd#1#2 {S_{ {\mbox{ {\tiny $\!\!\! #1 $} \rm}}}^{#2}}
\def \dWsp {\dsu{E}{F} (\Psi_{4} \sou{A} \sou{B} \sou{C} \sou{D} ) }
\def \P#1 {\Psi_{#1}}
\def \epsu#1 {\eps^{ {\mbox{ {\tiny $\!\!\! #1 $}\rm}}} }
\def \epsd#1 {\eps_{ {\mbox{ {\tiny $\!\!\! #1 $}\rm}}} }
\def \t#1 { \dot #1 }
\def \skd#1#2  {\delta^{ {\mbox{ {\tiny $\!\!\! #1 $} \rm}}}
                      _{ {\mbox{ {\tiny $\!\!\! #2 $} \rm}}} }
\def \kd#1#2 {\delta^{#1}_{#2}}
\def \sgu#1#2  {\sigma^{ {\mbox{ {\tiny $\!\!\! #1 $} \rm}}}
                    _{ #2 }}
\def \csgu#1#2  {{\bar \sigma}^{ {\mbox{ {\tiny $\!\!\! #1 $} \rm}}}
                    _{ #2 }}

\def \sgd#1#2  {\sigma_{ {\mbox{ {\tiny $\!\!\! #1 $} \rm}}}
                    ^{  #2 }}
\def \otud#1#2#3 { #1 ^{#2}_{\ #3}}
\def \otdu#1#2#3 { #1 _{#2}^{\ #3}}
\def \ospsmuu#1#2#3 {#1^{{\cal #2 }
                     { {\mbox{ {\tiny $\!\!\! #3 $} \rm}}}}}
\def \ospsmdd#1#2#3 {#1_{{\cal #2 }
                     { {\mbox{ {\tiny $\!\!\! #3 $} \rm}}}}}
\def \ospsmud#1#2#3 {#1^{\cal #2 }
                       _{\mbox{ {\tiny $ \ #3 $} \rm}}}
\def \ospsmdu#1#2#3 {#1_{\cal #2 }
                       ^{\mbox{ {\tiny $ \ #3 $} \rm}}}
\def \ospsmddu#1#2#3#4 {#1_{{\cal #2 }{\mbox{ {\tiny $ \!\! #3 $} \rm}}}
                       ^{\mbox{ {\tiny $ \ \ \ \ #4 $} \rm}}}
\def \nderPsi {\dsu{C_{n}}{X_{n}} \dots \dsu{C_{1}}{X_{1}}
(\Psi_{4} \sou{A} \sou{B} \sou{C} \sou{D} )  }
\def \nderPt {\dsu{C_{n}}{X_{n}} \dots \dsu{C_{1}}{X_{1}}
(\Psi_{3} \sou{(A} \sou{B} \sou{C} \siu{D)} )  }
\def \ospS {\ospu{S}{A_1 \dots A_m \dot X_1 \dots \dot X_k } }
\def \ospSvar {\ospu{S}{B_1 \dots B_m \dot Y_1 \dots \dot Y_k } }